ORIGINAL PAPER

# UVB radiation induced effects on cells studied by FTIR spectroscopy

**Lucia Di Giambattista · P. Grimaldi · S. Gaudenzi ·
D. Pozzi · M. Grandi · S. Morrone · I. Silvestri ·
A. Congiu Castellano**

**Abstract** We have made a preliminary analysis of the results about the effects on tumoral cell line (lymphoid T cell line Jurkat) induced by UVB radiation (dose of 310 mJ/cm$^2$) with and without a vegetable mixture. In the present study, we have used two techniques: Fourier transform infrared spectroscopy (FTIR) and flow cytometry. FTIR spectroscopy has the potential to provide the identification of the vibrational modes of some of the major compounds (lipid, proteins and nucleic acids) without being invasive in the biomaterials. The second technique has allowed us to perform measurements of cytotoxicity and to assess the percentage of apoptosis. We already studied the induction of apoptotic process in the same cell line by UVB radiation; in particular, we looked for correspondences and correlations between FTIR spetroscopy and flow cytometry data finding three highly probable spectroscopic markers of apoptosis (Pozzi et al. in Radiat Res 168:698–705, 2007). In the present work, the results have shown significant changes in the absorbance and spectral pattern in the wavenumber protein and nucleic acids regions after the treatments.

## Introduction

Infrared (IR) spectroscopy is a technique able to monitor the different phases of conformational and functional changes in complex biological samples, such as tissues or cell cultures. This technique allows to study the nature of the chemical bonds and their molecular environment in a given sample (Gaudenzi et al. 2004; Andrus and Strickland 1998). Our research is aimed to understand how the changes of IR spectra are correlated to different treatments, UVB radiation and a vegetable mixture, by using FTIR spectroscopy coordinated with flow cytometry and optical microscopy.

We have already made a similar study using only the UVB radiation on Jurkat cells (Pozzi et al. 2007), now we introduce a particular mixture to test a possible changes in the apoptotic process induced by UVB radiation. At present, the most medical studies are focused on control cell death because the loss of apototic process may be as important as increased cell growth in many tumors, so the medicinal natural research has introduced the use of natural compounds operating at molecular level (Chor et al. 2005).

We have chosen a natural mixture which is composed of some vegetables substances (*Epilobium, Urtica dioica*, and a lichen, *Evernia Prunastri*) in hydro-alcoholic solution.

L. Di Giambattista (✉) · P. Grimaldi · S. Gaudenzi ·
A. Congiu Castellano
Dipartimento di Fisica, Università di Roma Sapienza,
Piaz.le A. Moro, 2, 00185 Rome, Italy
e-mail: l.digiambattista@caspur.it

L. Di Giambattista
CISB (Interdepartmental Research Centre for Biomedical Systems Models and Information Analysis), Rome, Italy

D. Pozzi · S. Morrone · I. Silvestri
Dipartimento di Medicina Sperimentale,
Università di Roma Sapienza, V.le Regina Elena,
324, 00161 Rome, Italy

M. Grandi
Poliambulatorio La Torre, V. M. Ponzio,10,
10141 Torino, Italy

Lichens produce different secondary products (*lichen metabolites*) with filtering properties of UV radiation, in particular the phenolics metabolites that are partially secreted to lichen cortex and produced a protective screen against UV radiation (Legaz et al. 2001). By flow cytometry, we have analyzed the level of toxicity of the mixture (Tuschl and Schwab 2004) and we have established the percentage of apoptotic cells induced by UVB radiation with and without this mixture.

These our initial results have indicated that it is possible to relate the shift in the spectral signatures with the variations of the apoptotic process induced by the different treatments.

## Materials and method

### Cells and culture conditions

The Jurkat cells CD3+/CD2+, a lymphoid T cell line, were used for this study. Cells were cultured in RPMI 1640 medium added with 10% fetal bovine serum, 1% penicillin–streptomycin and 1% L-glutamine in humidified atmosphere (95% air) and 5% $CO_2$ at 37°C.

The unsynchronized cell population was used both for the untreated (control) and the treated sample at a routinely viability of 98% (determined by the Trypan blue exclusion test).

### Cell treatment

For our study, we used a particular mixture constituted from *Epilobium*, *Evernia Prunastri Arch* and *Urtica dioica*, in hydro-alcoholic solution.

The cellular samples were divided into four groups: cells irradiated with UVB, cells treated with UVB and some doses of the solution, cells treated with some doses of the solution and finally two control samples at 0 and 24 h after the beginning of treatment.

Preliminary test on the solution toxicity showed that doses higher than 30 µl cause a toxic effect on cells. Thus, we have treated the cellular sample with the following doses: 5, 10, 20, 30 µl.

At the same time, cellular samples treated only with 5, 10, 20, 30 µl of solution and samples exposed to UVB radiation with and without the solution were placed inside the incubator and cultured for 60, 210 and 360 min until the cytofluorimetric and spectroscopic measurements. The UVB exposure conditions have been the same used to induce the apoptosis (30 cm for 30 min).

### UVB radiation

A Philips TL20 W/12 lamp emitting 2.1 W at 310 nm was used to induce apoptosis through UVB irradiation (Novak et al. 2004). The exposure time of 30 min at a distance of 30 cm, corresponding to a dose of 310 mJ/cm² (confirmed by radiometer Gigahertz-Optik GmbH, Germany) produced the maximum percentage of apoptotic cells detected by flow cytometry.

Then, control and irradiated cells were cultured in the incubator for 60, 210, 360 min before the cytofluorimetric and spectroscopic measurements.

### Flow cytometry

Flow cytometry with double staining of Annexin V-FITC/ (PI) propidium iodide was performed to detect both apoptosis and necrosis. The percentage of apoptotic cells was determined by green fluorescence emitted by Annexin V-FITC bound to phosphatidylserine that is exposed to the outer leaflet of the membrane of apoptotic cells, while propidium iodide detects necrotic cells in the samples. Before the measurements, $1 \times 10^6$ cells were extensively washed in (PBS) Phosphate buffered saline, rinsed with Hepes buffer, resuspended in the same buffer, and incubated at room temperature for 5–15 min in the dark after the addition of Annexin V-FITC; just before the measurements, PI was added to the cell suspension.

The cytofluorimetric analysis was performed using a FACSCalibur (BD Biosciences, San Jose, CA) equipped with an argon-ion laser at an optimal excitation wavelength of 488 nm.

The PI intercalated into DNA, after spontaneously going into the cells, allow us to detect the toxicity threshold of mixture dose (Di Pietro et al. 2005). These measures were performed adding 10 µl of PI solution (10 mg/100 ml PBS) before cytometry detection.

### Optical microscopy

The morphological changes in cellular samples were detected through a Leica DMIL microscope equipped with a digital camera Olympus for the acquisition of frames. Cell images were acquired using a 40× magnification at 60, 210, 360 min after the treatments, placing the cells, with their culture medium, in uncovered culture plates.

### FTIR spectroscopy

Cell spectra were collected using an FTIR/410 Jasco Fourier Transform IR spectrometer equipped with an ATR-PRO410-S single reflection ATR accessory with a 45° single reflection ZnSe horizontal crystal plate.

Measurements were performed at room temperature using a conductive ceramic coil mounted in a water-cooled copper jacket source, a KBr beam splitter, an optical path purged continuously with gaseous nitrogen and a TGS detector.

**Table 1** Comparison of the percentage of viable cells at 0 and 24 h for different treatments (mixture and UVB)

| Treatment | Doses | % Viable cells | |
|---|---|---|---|
| | | 0 (h) | 24 (h) |
| Mixture (μl) | 5 | 89 | 87 |
| | 10 | 89 | 87 |
| | 20 | 89 | 88 |
| | 30 | 89 | 88 |
| UVB Radiation (mJ/cm²) | 310 | 89 | 67 |

For each spectrum, a resolution of 4 cm⁻¹ was used, 64 interferograms were coadded and apodized with a triangular function. ATR penetration depth compensation, was performed by using the Spectral Manager Analysis software.

Then, all IR spectra were studied in the spectral region from 1,760 to 900 cm⁻¹, baseline corrected and intensity normalized at 1,541 cm⁻¹; the experimental data were processed by using Microcal Origin 7.5 software. Three runs of FTIR, flow cytometry and microscopy experiments were performed.

## Results and discussion

### Flow cytometry

To quantify the toxicity of the treatments (mixture dose and UVB radiation), we have assessed the viability of cell population with PI: this marker is a DNA intercalating dye that is excluded by cells that have their plasma membrane integrity preserved (live cells), but it enters and stains DNA in cells that have damaged membranes, necrotic cells (Table 1).

The sample treated with mixture reveals no toxic effect up till 30 μl dose from 0 to 24 h, while the percentage of viable cells decreases at 24 h for UVB-treated sample. To assess the apoptosis, we have used FITC-coniugated Annexin V and PI.

In Fig. 1, we have reported the most significant results for treated cells with a dose of 30 μl (Van Engeland et al. 1998).

This analysis has shown how the mixture at different doses has modified the effect of UVB radiation between 0 and 6 h.

Finally, the experimental data were analyzed by repeated measures ANOVA. Through a *Fisher-Test* and the degrees of freedom ($F_{(3,12)}$) of Fisher distribution, we have evaluated and compared the FRatio ($R$) with FRatio* ($R*$) deducing by fixed significance level $\alpha$. The results show a significant difference between the treatments ($R* = 3.49$, $R = 10.15$ $p \leq 0.0013$ for live, $R = 11.67$ $p \leq 0.001$ for early apoptosis and $R = 4.45$ $p \leq 0.002$ for late apoptosis).

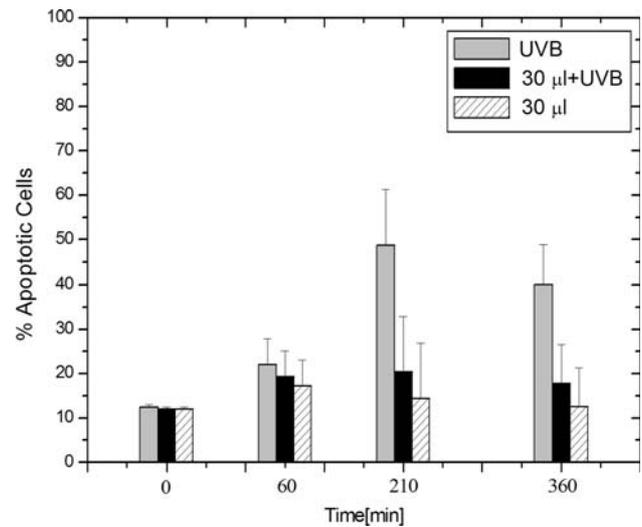

**Fig. 1** Comparison of the percentage of apoptotic cells as a function of time (0–6 h) for the samples treated with and without the UVB radiation

### Optical microscopy

To understand the effects of the treatments on the sample, we have observed the cellular morphology of the untreated and treated samples by the optical microscopy. The morphological changes are mainly due to UVB radiation. We have acquired the cellular images at 210 min after the treatment when we have observed the maximum of the early apoptotic process (Fig. 2).

As shown in Fig. 2 with white arrows, the apoptotic bodies are present only in the sample treated with UVB (Fig. 2b) and in the sample with mixture plus UVB (Fig. 2d) (Fadeel and Orrenius 2005).

### FTIR spectroscopy

In this preliminary study, we have shown the spectrum of the control sample only in the wavenumber ranges from 1,800 to 900 cm⁻¹ (Fig. 3); in Table 2, we have reported the assignments of commonly found absorption peaks in the IR spectra of cells (Jamin et al. 2003).

In particular, we have studied proteins (Amide I and Amide II bands centered at 1,646 cm⁻¹ and 1,541 cm⁻¹, respectively), and nucleic acids (1,300–1,000 cm⁻¹ and 1,000–950 cm⁻¹) regions. All FTIR spectra exhibited a good signal-to-noise ratio and were highly reproducible.

The following spectroscopic shift peaks were analyzed:

- $\alpha$-helix (a component of the band $A_1$)
- $P_3$ (asymmetric phosphate stretching vibrations)

**Fig. 2** The images of the cells at 210 min with ×40 magnifications: **a** untreated cells (control), **b** cells irradiated by UVB radiation, **c** cells treated with 30 μl of the mixture, **d** cells treated with 30 μl of the mixture plus UVB radiation. In **b–d**, the *arrows* indicate the apoptotic bodies which are absent in **a–c**

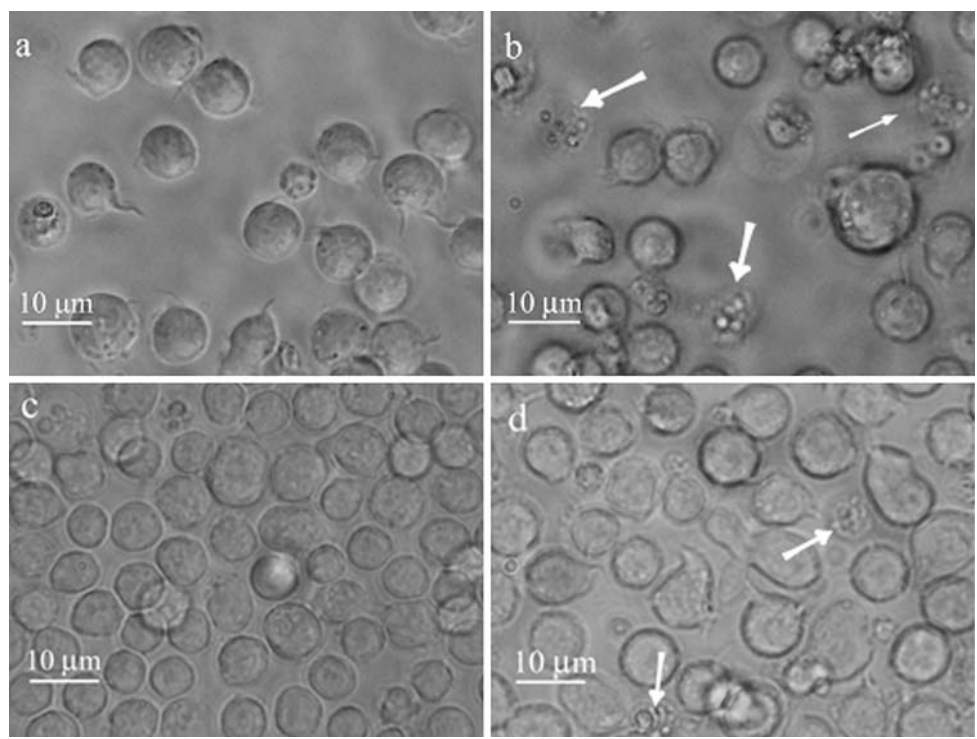

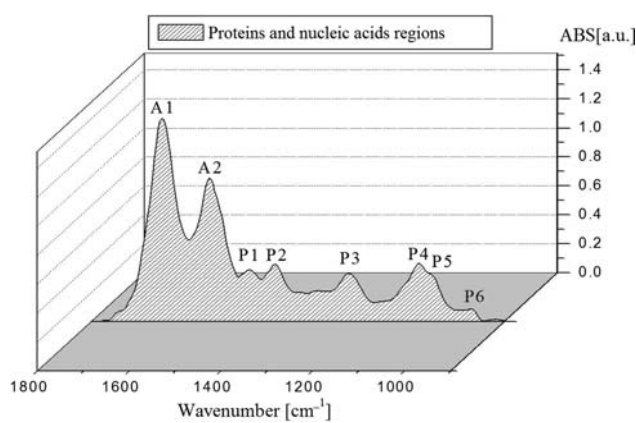

**Fig. 3** IR spectra of the control sample: proteins and nucleic acids regions

**Table 2** Representative frequencies and assignments of the major vibrational bands in FTIR spectra of cells

| Absorption bands (cm$^{-1}$) | Assignments |
| --- | --- |
| 1,646 (*A1*) | Amide I (–C=O stretching), proteins |
| 1,541 (*A2*) | Amide II (–N–H bending, –C–N stretching), proteins |
| 1,454 (*P1*) | –CH$_2$ scissoring/–CH$_3$ asymmetric bending, proteins, lipids |
| 1,399 (*P2*) | –COO$^-$ symmetric stretching, proteins, lipids |
| 1,244 (*P3*) | –PO$_2^-$ asymmetric stretching, nucleic acids, phospholipids |
| 1,085 (*P4*) | –PO$_2^-$ symmetric stretching, nucleic acids, phospholipids |
| 1,050 (*P5*) | –C–O– stretching, carbohydrates |
| 967 (*P6*) | –PO$_4^-$ symmetric stretching, nucleic acids |

## Proteins and nucleic acids

The region from 1,800 to 1,480 cm$^{-1}$ is dominated by the absorption modes of Amide I (*A1*) centered at 1,646 cm$^{-1}$ and Amide II (*A2*) at 1,541 cm$^{-1}$ which originate from the vibrations of the amide groups (CO–NH). The shape of the Amide I band is influenced by the overlap of secondary structure of proteins: these components are α-helix between 1,645 and 1,666 cm$^{-1}$, β-sheets between 1,613 and 1,637 cm$^{-1}$, β-turns between 1,666 and 1,682 cm$^{-1}$ and random coil between 1,637 and 1,645 cm$^{-1}$. The behavior of the α-helix band (1,664 cm$^{-1}$) seems to be particularly significant in the cells treated with mixture and UVB radiation (Gault and Lefaix 2003).

As shown in Fig. 4, it is possible to observe a correspondence between the shift of this spectral component and the apoptotic process induced by UVB radiation in presence of the mixture (Hagenhofer et al. 1998). The correlation between the α-helix shift and the percentage of apoptotic cells shows that the resulting Pearson's correlation coefficients are 0.95 for UVB-treated sample and 0.91 for mixture + UVB-treated sample (*R*-value), indicating a probability for both samples of about 99% (*Fisher-Test*) for the significance of the linear regression. So, we have

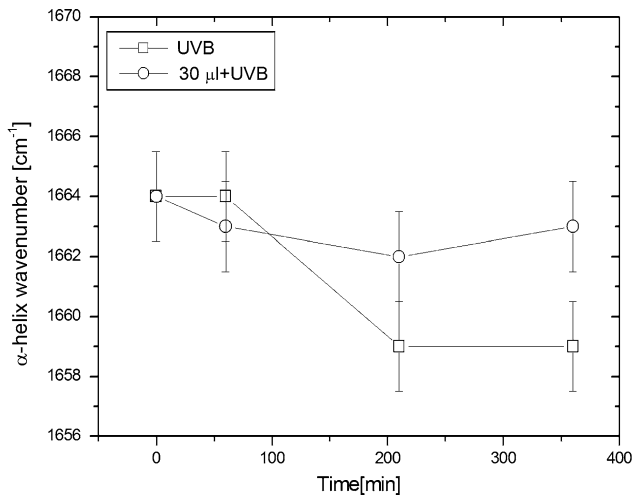

**Fig. 4** Shift of the α-helix spectral component for the followings samples: UVB, mixture plus UVB

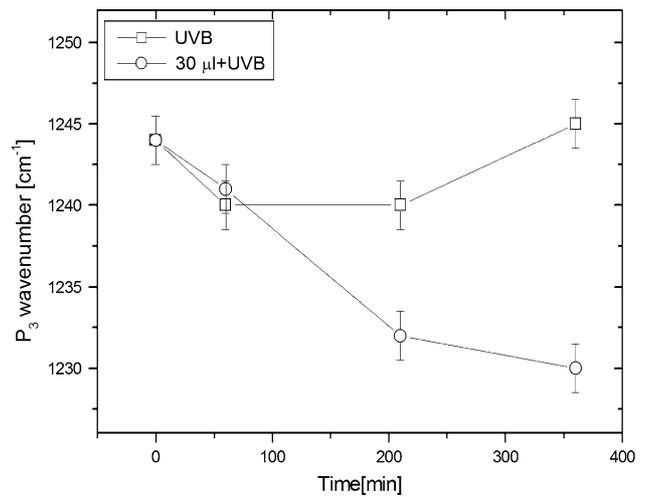

**Fig. 6** Shift of the spectral band of RNA/phosphodiester groups

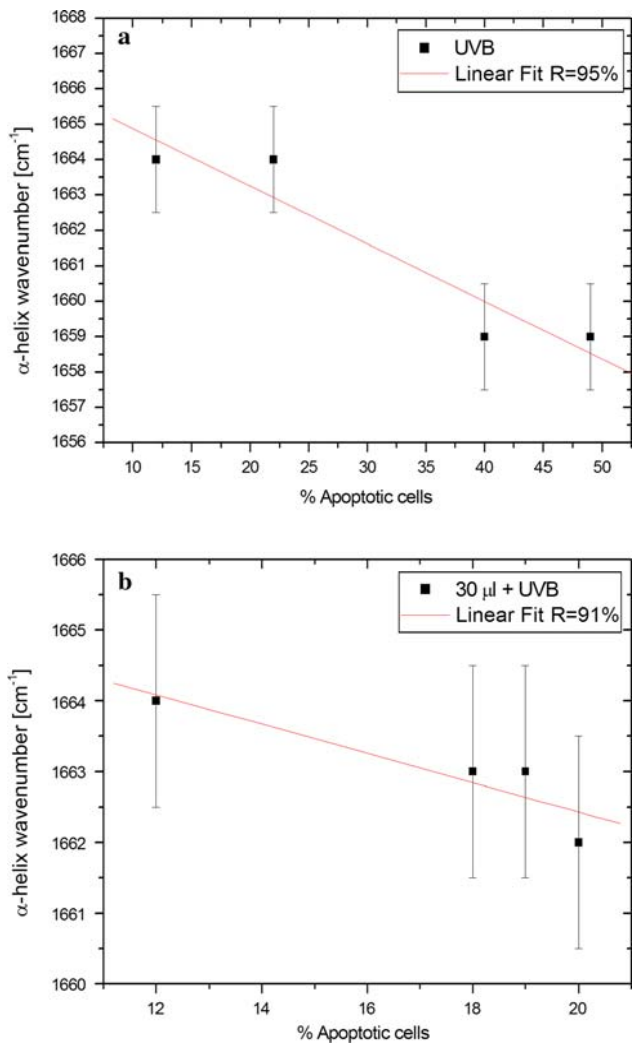

**Fig. 5 a–b** Linear correlation between the α-helix shift and the percentage of apoptotic cells believed that the shift of the α-helix can be defined as a "spectroscopic biomarker" of the increase of structural disorder induced by UVB radiation that the mixture can only delay (Fig. 5).

In the region of nucleic acids ($1,300$–$900 \, cm^{-1}$) significant changes were observed in the cellular sample in correspondence of the band centered at $1,244 \, cm^{-1}$ which originates primarily from the nucleic acid phosphodiester groups of the RNA (Fig. 6) (Ravant et al. 2001).

The wavenumber of this spectral band has undergone change between 0 and 360 min, in particular: at 210 min we have observed a significant shift of about $13 \, cm^{-1}$ to asymmetric $PO_2^-$ stretching of the DNA for a sample treated with the mixture plus UVB radiation which has shown a value of the percentage of apoptotic cells lower than the value of the cells treated with UVB radiation.

At the time of 360 min when we observe the reduction of the percentage of apoptotic cells, in UVB treated sample the $P_3$ shift is lost while in the presence of the mixture we observe or a permanent shift or a delay of the effect of the UVB radiation.

## Conclusions

In the present work, we have studied the behavior of the Jurkat cells treated with a vegetable mixture and UVB radiation. We have evaluated a biomarker (α-helix spectral component) derived from FTIR spectroscopic measurements which could be suitable for monitoring the degree of molecular disorder induced by UVB radiation. The observed spectral features are important for an understanding of the spectral differences between the cellular sample treated with and without the mixture. We underline that these results concern the experiment on cells in vitro.

Therefore, further measurements are in progress in order to better investigate the spectroscopic behaviours emerged in the current analysis.

**Acknowledgments** We are grateful to Prof. M. Severini for the supply of the UVB radiation source and to Dr. Stefano Belardinelli for his assistance in the laboratory experiments.